\documentclass[aps,prl,twocolumn,superscriptaddress,showpacs]{revtex4-1}
\usepackage{amsmath,amssymb,graphics,epsfig,epstopdf,color,verbatim,ulem,braket,tabularx}
\usepackage[colorlinks,linkcolor=blue,citecolor=blue,urlcolor=blue]{hyperref}

\graphicspath{{figures/}}

\begin{document}

\title{Visualizing a Bosonic Symmetry Protected Topological Phase \\ in an Interacting Fermion Model}
\author{Han-Qing Wu}
\author{Yuan-Yao He}
\address{Department of Physics, Renmin University of China, Beijing 100872, China}
\author{Yi-Zhuang You}
\affiliation{Department of Physics, University of California,
Santa Barbara, California 93106, USA}
\author{Tsuneya Yoshida}
\affiliation{Department of Physics, Kyoto University, Kyoto 606-8502, Japan}
\author{Norio Kawakami}
\affiliation{Department of Physics, Kyoto University, Kyoto 606-8502, Japan}
\author{Cenke Xu}
\affiliation{Department of Physics, University of California,
Santa Barbara, California 93106, USA}
\author{Zi Yang Meng}
\affiliation{Beijing National Laboratory
for Condensed Matter Physics, and Institute of Physics, Chinese
Academy of Sciences, Beijing 100190, China}
\author{Zhong-Yi Lu}
\address{Department of Physics, Renmin University of China, Beijing 100872, China}

\begin{abstract}

Symmetry protected topological (SPT) phases in free fermion and
interacting bosonic systems have been classified, but the physical
phenomena of interacting fermionic SPT phases have not been fully
explored. Here, employing large-scale quantum Monte Carlo
simulation, we investigate the edge physics of a bilayer
Kane-Mele-Hubbard model with zigzag ribbon geometry. Our unbiased
numerical results show that the fermion edge modes are gapped out
by interaction, while the bosonic edge modes remain gapless at the
$(1+1)d$ boundary, before the bulk quantum phase transition to a
topologically trivial phase. Therefore, finite fermion gaps both
in the bulk and on the edge, together with the robust gapless
bosonic edge modes, prove that our system becomes an emergent
bosonic SPT phase at low energy, which is directly observed in an interacting fermion lattice model.

\end{abstract}


\date{\today} \maketitle

\textit{Introduction.} Symmetry protected topological (SPT) phases
are bulk gapped states with either gapless or degenerate edge
excitations protected by symmetries. The SPT phases in free
fermion systems, like topological
insulators~\cite{kane2005a,kane2005b,fukane,moorebalents2007,roy2007},
acquire metallic edge states and have been fully
classified~\cite{ludwigclass1,kitaevclass}. On the other hand,
although bosonic SPT phases have been formally classified and
constructed as well from group cohomology~\cite{Chen2012,Chen2013}
and field
theories~\cite{luashvin,senthilashvin,xusenthil,xuclass}, there
has been little study about realization of bosonic SPT states in
condensed matter systems, except for the well-known one-dimensional Haldane
phase that is realized in a spin-1 Heisenberg
model~\cite{haldane1,haldane2} and some proposals of realizing a
two-dimensional bosonic SPT state in cold atom systems~\cite{levinsenthil}.
Using the same ``flux-attachment" picture as
Ref.~\onlinecite{levinsenthil}, lattice models of bosonic integer
quantum Hall states have been
studied~\cite{Furukawa2013,YinChen2015,Sterdyniak2015,Yohei2016,TianSheng2016}.

\begin{figure}[htp!]
\centering
\includegraphics[width=0.85\columnwidth]{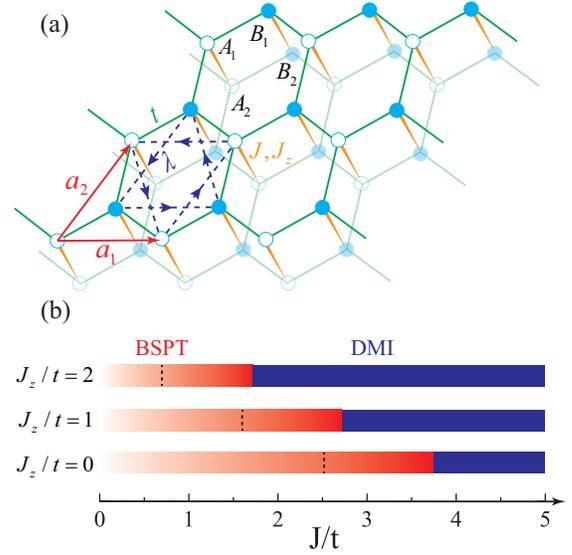}
\caption{(Color online) (a) Illustration of AA-stacked honeycomb
ribbon ($L_{\mathbf{a}_{1}}=3$, $L_{\mathbf{a}_{2}}=3$) with
periodic (open) boundary condition along the $\mathbf{a}_{1}$
($\mathbf{a}_{2}$) direction. $\mathbf{a}_{1}=(1, 0)$ and
$\mathbf{a}_{2}=(1/2, \sqrt{3}/2)$ are the primitive translation
vectors. $A_{1}$, $B_{1}$, $A_{2}$ and $B_{2}$ are the four
sublattices within one unit cell. (b) $J$-$J_z$ phase diagram of
bilayer Kane-Mele-Hubbard model. The bosonic SPT (BSPT, red) and
dimer Mott insulator (DMI, blue) phases are separated by a bulk
transition. The dashed lines inside BSPT denote the $J$ values,
above which one can clearly see the exponential decay of the
single-particle Green's function at the boundary from our
finite-size calculations. The relative range of such region
becomes wider as $J_z$ increases.} \label{fig:BKMH}
\end{figure}

Recently it was proposed that instead of directly studying bosonic
systems, the physics of bosonic SPT states can be mimicked by
interacting fermionic systems, in the sense that its low energy
physics is completely identical to bosonic SPT states~\cite{xufb}.
For example, in an interacting fermion model on the AA-stacked
bilayer Kane-Mele-Hubbard model, a {\it bona fide}
interaction-driven topological phase transition has been studied
in our previous papers~\cite{Kevin2015,YYHe2016a,YYHe2016b}. A
direct continuous quantum phase transition between a quantum spin
Hall (QSH) phase and a topologically trivial Mott insulator was
found via large-scale quantum Monte Carlo (QMC) simulations. At
the critical point, only the bosonic spin and charge gaps are
closed, while the bulk single-particle excitations remain open.
This transition can be described by a $(2+1)d$ $O(4)$ nonlinear
sigma model with a topological
$\Theta$-term~\cite{CKXu2013,Kevin2015,YYHe2016a}. However, as for
the physics on the edge, although the field theory and
renormalization group analysis~\cite{YZYou2016} provide us with
analytical evidence of a gapless bosonic edge, which
is supported by an extended version of dynamical mean-field theory
calculation at finite temperatures\cite{Yoshida2016}, unbiased
numerical evidence that can prove the conclusion is still
demanded, and it is the task of this paper.

Here, we employ large-scale QMC simulation to the zigzag ribbon
geometry, i.e., the bilayer Kane-Mele-Hubbard model with periodic
boundary condition along the $\mathbf{a}_1$ direction and open
boundary along the $\mathbf{a}_2$ direction [see Fig.~\ref{fig:BKMH}
(a)]. On finite-size ribbon, our unbiased results unveil a
substantial region ($\sim t$) of bosonic SPT phase from the
exponential decay of the single-particle Green's function along
the boundary before the bulk quantum phase transition, while the
gapless $O(4)$ bosonic modes prevail on the edge with power-law
correlation functions.

\textit{Model and method.} The
Hamiltonian~\cite{YYHe2016a,YZYou2016} of the AA-stacked bilayer
Kane-Mele-Hubbard model is given by
\begin{equation}
\begin{split}
\hat{H}&=-t\sum_{\xi\langle i,j\rangle\alpha}(\hat{c}_{\xi i\alpha}^{\dagger}\hat{c}_{\xi j\alpha}+\hat{c}_{\xi j\alpha}^{\dagger}\hat{c}_{\xi i\alpha})\\
&+i\lambda\sum_{\xi\langle\langle i,j \rangle\rangle\alpha\beta}\nu_{ij}(\hat{c}_{\xi i\alpha}^{\dagger}\sigma_{\alpha\beta}^{z}\hat{c}_{\xi j\beta}-\hat{c}_{\xi j\beta}^{\dagger}\sigma_{\beta\alpha}^{z}\hat{c}_{\xi i\alpha}) \\
&-\frac{J}{8}\sum_{i}\left[(\hat{D}_{1i,2i}+\hat{D}_{1i,2i}^{\dagger})^{2}-(\hat{D}_{1i,2i}-\hat{D}_{1i,2i}^{\dagger})^{2}\right] \\
&-\frac{J_{z}}{4}\sum_{i}\left[(\hat{n}_{1i\uparrow}-\hat{n}_{1i\downarrow})-(\hat{n}_{2i\uparrow}-\hat{n}_{2i\downarrow})\right]^{2},
\label{eq:BilayerHmlt}
\end{split}
\end{equation}
with
$\hat{D}_{1i,2i}=\sum_{\sigma}\hat{c}_{1i\sigma}^{\dagger}\hat{c}_{2i\sigma}$.
Here $\alpha$, $\beta$ denote the spin species and $\xi=1,2$ stand
for the layer index. The first term in Eq.~(\ref{eq:BilayerHmlt})
describes the nearest-neighbor hopping [green lines in
Fig.~\ref{fig:BKMH} (a)] and the second term represents
spin-orbital coupling $\lambda/t=0.2$ [blue lines with arrows in
Fig.~\ref{fig:BKMH} (a)]. The third term $J$ is the interlayer
antiferromagnetic Heisenberg (approximated)
interaction~\cite{YYHe2016a}, and the last term $J_{z}$ denotes
the interlayer antiferromagnetic Ising (approximated)
interaction~\cite{YZYou2016}. When $J/t>0$ and $J_{z}/t>0$, we can
prove that there is no fermion sign problem in the QMC
calculations~\cite{YZYou2016}.

This Hamiltonian possesses a high symmetry, $SO(4)\times
SO(3)$~\cite{YYHe2016a,YZYou2016}. When $J_{z}/t=0$, in the bulk,
$J$ drives a continuous quantum phase transition from a QSH phase to an interlayer dimer phase at
$J_{c}/t\approx 3.73$, and since there is no spontaneous symmetry
breaking at both sides of this transition, it is dubbed as a
{\it{bona fide}} interaction-driven topological phase
transition~\cite{YYHe2016a}. On the other hand, when $J/t=0$, it is
perceivable that $J_{z}$ will eventually drive the system into a
spin-density-wave phase with magnetization along the $z$ direction
(SDW-Z) which spontaneously breaks the $SO(3)$ symmetry and
time-reversal symmetry. Our numerical data shows that the SDW-Z
order establishes when $J_z/t > 2$. More information about the
$J-J_{z}$ phase diagram is given in the Supplemental
Material~\cite{Suppl}.

The QSH phase still survives when the interlayer interactions are
not sufficiently strong. However, we will show that the gapless
edge modes in the interacting QSH phase are carried by bosons
emerging from interacting fermionic degrees of freedom, hence the
system is actually in a bosonic SPT state before the bulk phase
transition [the BSPT phase in Fig.~\ref{fig:BKMH} (b)]. This
conclusion is drawn upon the numerical observation of exponential
decay of a single-particle Green's function on the edge before the
bulk quantum phase transition, while at the same time bosonic
$O(4)$ correlation functions present a clear power-law decay.

The QMC method employed here is the projective auxiliary-field
quantum Monte Carlo approach~\cite{AssaadEvertz2008,Meng2010}. It
is a zero-temperature version of the determinantal QMC algorithm.
The specific implementation of the QMC method on the model in
Eq.~(\ref{eq:BilayerHmlt}) is presented in Ref.~\cite{YYHe2016a}.
The projection parameter is chosen at $\Theta=50/t$ and the
Trotter slice $\Delta\tau=0.05/t$. Since the gapless edge modes
are hallmarks of SPTs, we perform the simulation with periodic
(open) boundary condition along the $\mathbf{a}_{1}$
($\mathbf{a}_{2}$) direction [see Fig.~\ref{fig:BKMH} (a)]. The
main results in this paper are obtained from a ribbon with
$L_{\mathbf{a}_{1}}=27, L_{\mathbf{a}_{2}}=9$ which is large
enough to obtain controlled representation of thermodynamic limit
behaviors of the BSPT phase in Fig.~\ref{fig:BKMH} (b).

\begin{figure}[tp!]
  \centering
  \includegraphics[width=0.8\columnwidth]{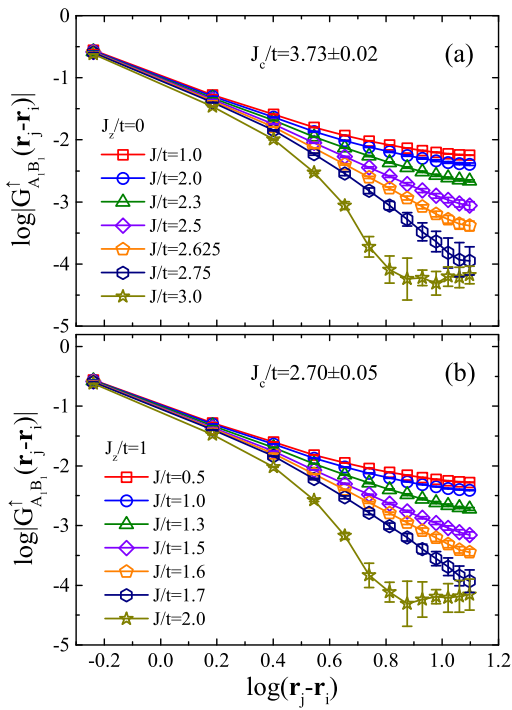}
\caption{(Color online) The log-log plot of single-particle
Green's function at the boundary as a function of interlayer
antiferromagnetic interaction $J/t$ when (a) $J_{z}/t=0$ and (b)
$J_{z}/t=1$. In both cases, results show the exponential decay
before the bulk topological phase transition $J_{c}/t$.}
  \label{fig:OccEST}
\end{figure}

\textit{Edge analysis.} In the noninteracting limit, the bilayer
Kane-Mele model supports four fermionic edge modes: two
left-moving up-spin modes and two right-moving down-spin modes
from both layers, respectively. They are denoted by the boundary
fermion fields $c_{\xi\alpha}$ ($\xi=1,2$,
$\alpha=\uparrow,\downarrow$). Following the standard Abelian
bosonization procedure, we can rewrite
$c_{\xi\alpha}=\kappa_{\xi\alpha}e^{i\phi_{\xi\alpha}}/\sqrt{2\pi
a}$, where $a$ is a short distance cutoff and
$\kappa_{\xi\alpha}$ is the Klein factor that ensures the
anticommutation of the fermion operators. As we turn on the
interaction, in terms of the bosonized degrees of freedom
$\phi=(\phi_{1\uparrow},\phi_{2\uparrow},\phi_{1\downarrow},\phi_{2\downarrow})$,
the effective action for the interacting edge modes reads
\begin{equation}
\begin{split}
S&=\int d\tau dx\frac{1}{4\pi}(\partial_x\phi^\intercal K \partial_\tau\phi+\partial_x\phi^\intercal V \partial_x \phi)-\lambda\cos(l_0^\intercal\phi),\\
K&=\left(\begin{smallmatrix}1&&&\\&1&&\\&&-1&\\&&&-1\end{smallmatrix}\right),
V=v_0\left(\begin{smallmatrix}1&u&-g&g\\u&1&g&-g\\-g&g&1&u\\g&-g&u&1\end{smallmatrix}\right),
\end{split}
\end{equation}
where $g=J_z/(4\pi v_0-J_z)$, $u=(J_z+J)/(4\pi v_0-J_z)$ and $v_0$
is the bare velocity of the edge modes. $\lambda\propto J$ is the
backscattering term induced by the interlayer Heisenberg
interaction with the corresponding charge vector
$l_0=(1,-1,-1,1)^\intercal$. The scaling dimension of
$\cos(l_0^\intercal\phi)$ is
\begin{equation}\label{eq: Delta0}
\Delta_0=\frac{2(1-u-2g)}{\sqrt{(1-u)^2-4g^2}}.
\end{equation}
Without the Ising interaction $J_z$ (i.e.~$g\to0$), the operator
$\cos(l_0^\intercal\phi)$ is marginal from the scaling dimension
$\Delta_0=2$. Further renormalization group (RG)
analysis\cite{YZYou2016} shows that the term
$\lambda\cos(l_0^\intercal\phi)$ is marginally relevant, meaning
that the fermionic edge modes of the non-interacting QSH state are
unstable to the interaction $J$. As long as $J$ is turned on, the
boundary fermions will be gapped out by the interaction, leaving
only bosonic edge modes described by the spin
$c_{1\uparrow}^\dagger c_{1\downarrow}-c_{2\uparrow}^\dagger
c_{2\downarrow}$ and charge $c_{1\uparrow}
c_{2\downarrow}-c_{1\downarrow} c_{2\uparrow}$ fluctuations.
However, due to the marginal nature of RG flow, the boundary
fermion gap could be very small for small $J$, which is hard to
resolve in our finite-size numerical study. The positive $J_z$
interaction (i.e.~$g>0$) helps to boost the RG flow by reducing
the scaling dimension $\Delta_0$ according to Eq.~\eqref{eq:
Delta0}, such that $J$ becomes relevant and the gap in the
single-particle (fermionic) spectrum can be observed in numerics
for smaller $J$ as well. In the following, we will show that with
moderate interaction $J$, the QSH edge modes indeed become
bosonic at low energy, resembling the key feature of BSPT states.
The interaction $J_z$ will help to enhance the fermion gap and
make the BSPT edge modes more prominent in a finite-size system.

\textit{Numerical results.} Figures~\ref{fig:OccEST} (a) and (b)
show the single-particle Green's function
$G_{ij}^{\sigma}=\bra{\Psi}\hat{c}_{i\sigma}^{\dagger}\hat{c}_{j\sigma}\ket{\Psi}/\braket{\Psi|\Psi}$
along the edge as a function of $J/t$, at $J_z/t=0$ and $1$,
respectively. $\ket{\Psi}\propto
e^{-\Theta\hat{H}/2}\ket{\Psi_{T}}$ is the ground state wave
function projected from a trial wave function $\ket{\Psi_{T}}$~\cite{YYHe2016a}. We
see a clear exponential decay before the bulk transition at
$J_c/t\approx 3.73$ (for $J_z/t=0$) and $J_c/t\approx 2.7$ (for $J_z/t=1$). The
exponential decay of edge single-particle Green's function at
$J<J_c$ indicates that fermions are no longer gapless at the
boundary between our model system and a topologically trivial one (such as vacuum).

\begin{figure}
  \centering
  \includegraphics[width=0.8\columnwidth]{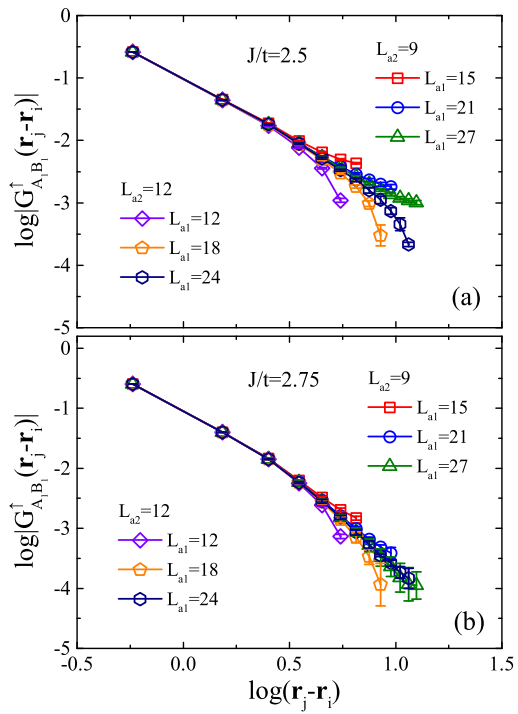}
\caption{Illustration of finite-size effects in the
single-particle Green's function along the edge for different
$L_{\mathbf{a}_1}$ and $L_{\mathbf{a}_2}$. (a) at $J/t=2.5,
J_{z}/t=0$, the exponential decay of the single-particle Green's
function acquires strong finite-size effect. (b) at $J/t=2.75,
J_{z}/t=0$, the finite size effect is absent and exponential decay
is seen for the chosen $L_{\mathbf{a}_1}$ and $L_{\mathbf{a}_2}$.
}
  \label{fig:OccFSE}
\end{figure}

To rule out the possible finite-size effect, we employ several
different ribbon geometries in the QMC calculations. From
Fig.~\ref{fig:OccFSE} (a), it is hard to determine whether the edge
single-particle Green's function will exponentially decay in the
thermodynamic limit when $J/t=2.5, J_{z}/t=0$ because of the
strong finite-size effect. However, when $J/t=2.75, J_{z}/t=0$, we
see a clear exponential decay no matter if $L_{\mathbf{a}_{1}}$ and
$L_{\mathbf{a}_{2}}$ are even or odd, large or small, and the single-particle
Green's function has a clear trend to truly exponential decay in
the thermodynamic limit.

\begin{figure}
  \centering
  \includegraphics[width=0.8\columnwidth]{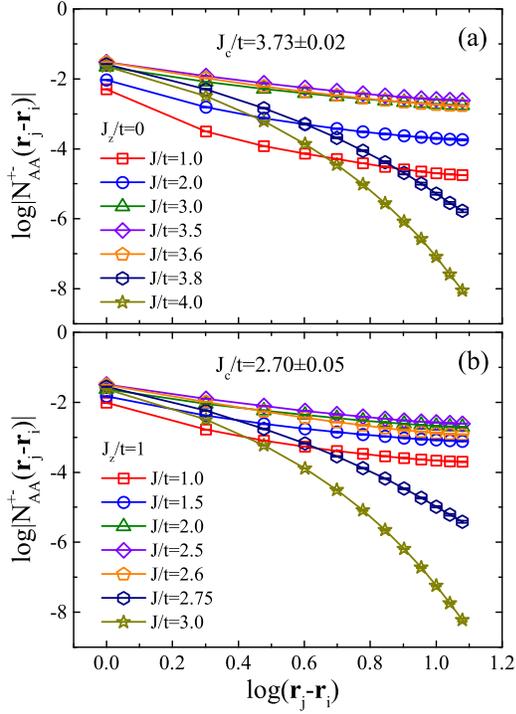}
\caption{(Color online) The log-log plot of equal-time
two-particle $O(4)$ vector correlation function at the boundary
for (a) $J_{z}/t=0$ and (b) $J_{z}/t=1$. Both panels show the
power-law decay behaviors before the bulk topological phase
transition at $J_{c}/t$.}
  \label{fig:SpmEST}
\end{figure}

The exponential decay of single-particle Green's function at the
boundary in the thermodynamic limit indicates that the gapless
fermion edge mode in the non-interacting case is gapped out by the
interlayer exchange interaction. Hence the fermion excitations
have a gap both in the bulk and on the edge~\cite{YYHe2016a}.
However, as shown in our edge analysis, the system can still be
non-trivial in the bosonic sector~\cite{YZYou2016}. To see this,
we calculate the XY spin (SDW-XY) correlation function and superconducting
pairing (SC) correlation function at the boundary. According to the analysis
in Ref.~\cite{YZYou2016}, we define them as
\begin{eqnarray}
N_{AA}^{+-}(\mathbf{r}_{j}-\mathbf{r}_{i})&=&\frac{1}{2}[S_{A_{1}A_{1}}^{\pm}(\mathbf{r}_{j}-\mathbf{r}_{i})-S_{A_{1}A_{2}}^{\pm}(\mathbf{r}_{j}-\mathbf{r}_{i})\nonumber\\
&-&S_{A_{2}A_{1}}^{\pm}(\mathbf{r}_{j}-\mathbf{r}_{i})+S_{A_{2}A_{2}}^{\pm}(\mathbf{r}_{j}-\mathbf{r}_{i})]\nonumber\\
\Delta_{AA}(\mathbf{r}_{j}-\mathbf{r}_{i})&=&\bra{\Psi}\hat{\Delta}_{iA_{1}A_{2}}^{\dagger}\hat{\Delta}_{jA_{1}A_{2}}\ket{\Psi}/\braket{\Psi|\Psi}
\label{eq:EdgeGF}
\end{eqnarray}
where
$S_{mn}^{\pm}(\mathbf{r}_{j}-\mathbf{r}_{i})=\bra{\Psi}\frac{1}{2}
(\hat{S}_{i}^{+}\hat{S}_{j}^{-}+\hat{S}_{i}^{-}\hat{S}_{j}^{+})\ket{\Psi}/\braket{\Psi|\Psi}$,
$m, n=A_{1},A_{2}$ denote the $A$ sublattice sites in the first
and second layer. $i$ and $j$ label the unit cells.
$\hat{S}^{+}_{i}$ is the spin flip operator and
$\hat{\Delta}_{iA_{1}A_{2}}^{\dagger}$ is the interlayer singlet
creation operator. Figures~\ref{fig:SpmEST} (a) and (b) show the
SDW-XY correlation function at the boundary as a function of
$J/t$. Before the bulk quantum phase transition, they all show the
power-law decay at $J<J_c$. Due to the $SO(4)$ symmetry, the
SDW-XY and SC correlation functions are exactly the same because
they rotate into each other~\cite{YYHe2016a,YZYou2016}. So the
physical bosonic boundary modes are simply the SDW-XY and SC
fluctuations on the boundary.

\begin{figure}
  \centering
  \includegraphics[width=0.8\columnwidth]{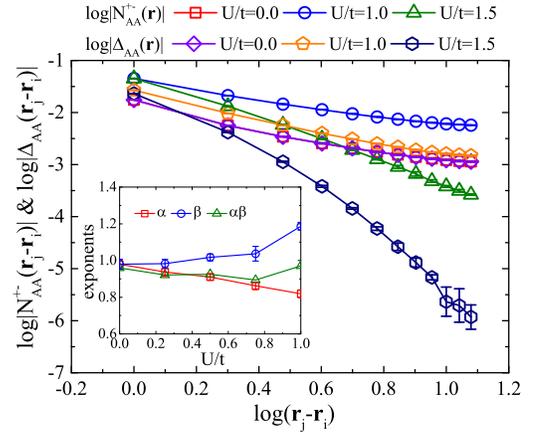}
\caption{(Color online) Edge spin $N^{+-}_{AA}(\mathbf{r})$ and
pairing $\Delta_{A_{1}A_{2}}(\mathbf{r})$ correlation functions
for increasing $U/t$, at $J/t=2.75$ and $J_{z}/t=0$. The inset shows the
extracted Luttinger parameters as a function of $U/t$.}
  \label{fig:luttinger}
\end{figure}

Turning on an extra on-site Hubbard interaction
$U\sum_{i}(\hat{n}_{i\uparrow}+\hat{n}_{i\downarrow}-1)^{2}$
(see Sec. VII in the Supplemental Material~\cite{Suppl} for the
$U/t$ path chosen in the bulk phase diagram) to our original model
Eq.~(\ref{eq:BilayerHmlt}) would break the $O(4)$ symmetry, and
change the scaling dimension of the spin and Cooper pair
operators. According to the bosonization analysis in
Ref.~\cite{YZYou2016}, the spin and pairing $O(4)$ bosonic modes
always have power-law correlation, with $N^{+-}_{AA}(\mathbf{r})
\propto |\mathbf{r}|^{-\alpha}$ and $\Delta_{AA}(\mathbf{r})
\propto |\mathbf{r}|^{-\beta}$. $\alpha$ and $\beta$ depend on the
Luttinger parameters, but their product remains a universal
constant: $\alpha\beta = 1$. This is due to the fact that, spin
and charge are a pair of conjugate variables at the boundary,
which is a physical consequence of the SPT state in the bulk. This
prediction is confirmed in our simulation. In
Fig.~\ref{fig:luttinger}, at $J/t=2.75$ $J_{z}/t=0$ and gradually
increasing $U/t$, $N^{+-}_{AA}(\mathbf{r})$ and
$\Delta_{AA}(\mathbf{r})$ have the same power law
$\alpha=\beta\sim1$ at $U/t=0$, but as $U/t$ increases, $\alpha$ and
$\beta$ start to deviate, but their product $\alpha\beta$ remains
close to $1$, as shown in the inset of Fig.~\ref{fig:luttinger},
until the bulk transition to a SDW-XY phase at $U_{c}/t\sim
1.3$~\cite{YYHe2016a,Suppl}.

\textit{Discussion.} In this paper, we have performed QMC
simulation for a proposed interacting lattice fermion model, and
explicitly demonstrated that this system shows a bosonic SPT
state, in the sense that the boundary has gapless bosonic modes,
but no gapless fermionic modes under interaction. Recently it was
also proposed that the same physics can be realized in an AB
stacking bilayer graphene under a strong out-of-plane magnetic
field and Coulomb interaction~\cite{ZhenBi2016}. Our model, though
technically different, should belong to the same topological
class, and it has the advantage of being sign problem free for QMC
simulation. Unbiased information of such a strongly correlated
system, including transport and spectral properties, can be
obtained from QMC simulation, and quantitative comparison with the
up-coming experiments is hence made possible.

\begin{acknowledgments}
The numerical calculations were carried out at the National Supercomputer Center in Guangzhou on the Tianhe-2 platform. Z.Y.M acknowledges the support from the Ministry of Science and Technology (MOST) of China under Grant No. 2016YFA0300502, the National Natural Science Foundation of China (NSFC Grants No. 11421092 and No. 11574359), as well as the National Thousand-Young-Talents Program of China. C.X. and Y.Z.Y. are supported by the David and Lucile Packard Foundation and NSF Grant No. DMR-1151208. T.Y. and N.K. are supported by JSPS KAKENHI No. 15H05855. H.Q.W., Y.Y.H., and Z.Y.L. acknowledge support from the NSFC Grants No. 11474356 and No. 91421304 and Special Program for Applied Research on Super Computation of the NSFC-Guangdong Joint Fund (the second phase).
\end{acknowledgments}

\bibliography{BilayerBib}

\clearpage
\widetext
\begin{center}
\textbf{\large Supplemental material: Visualizing a Bosonic
Symmetry Protected Topological Phase in an Interacting Fermion
Model}
\end{center}
\setcounter{equation}{0}
\setcounter{figure}{0}
\setcounter{table}{0}
\setcounter{page}{1}
\makeatletter
\renewcommand{\theequation}{S\arabic{equation}}
\renewcommand{\thefigure}{S\arabic{figure}}
\renewcommand{\bibnumfmt}[1]{[S#1]}

\section{I.$J_{z}/t=2$ results}

\begin{figure}[htp!]
  \centering
  \includegraphics[width=0.8\columnwidth]{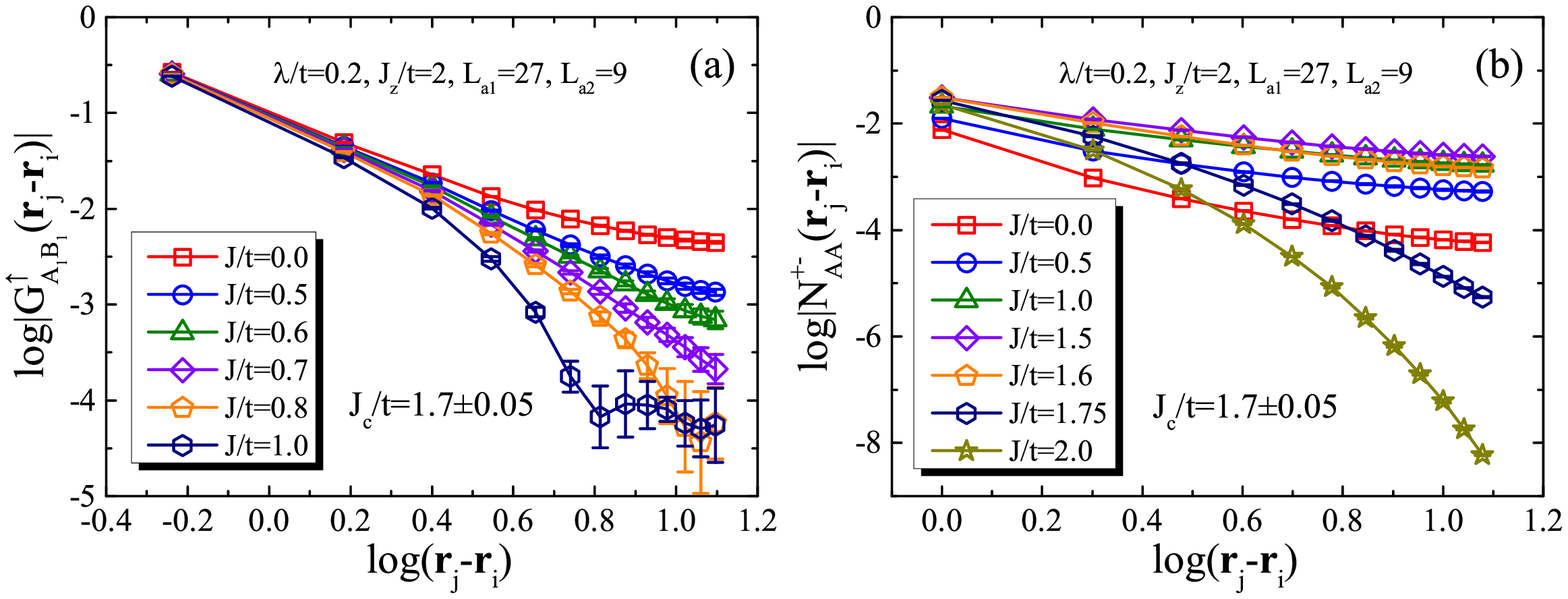}
\caption{Single-particle Green's function (a) and SDW-XY
correlation function (b) at the ribbon edge as a function of $J/t$
when $J_{z}/t=2$.}
  \label{fig:Jz2p0OccET}
\end{figure}

Fig.~\ref{fig:Jz2p0OccET} shows the single-particle Green's
function and SDW-XY correlation function at the ribbon edge as a
function of interlayer $J/t$ interaction when $J_{z}/t=2$. The bulk
quantum critical point is obtained from energy curves and SDW-XY
magnetic structure factors which will be shown in the following
section. The $J_{z}/t=2$ case shares the similar behavior as the
$J_{z}/t=0$ and $1$ cases. The single-particle Green's function at
the ribbon edge shows the exponential decay before the bulk
quantum phase transition, while the SDW-XY correlation function
still decays as a power-law behavior.

\section{II. magnetic orders}

\begin{figure}[htp!]
  \centering
\includegraphics[width=0.7\columnwidth]{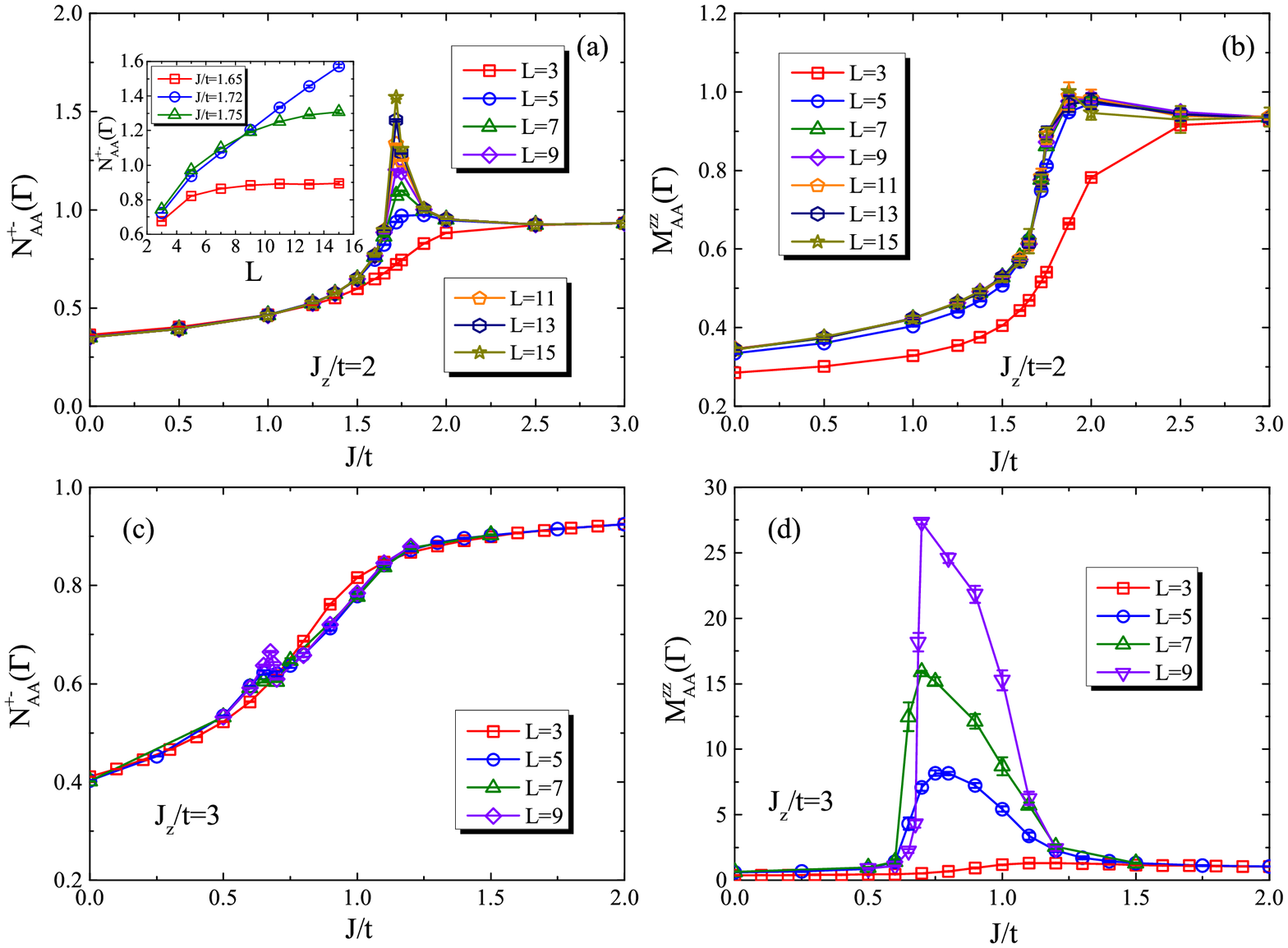}
\caption{SDW-XY (a,c) and SDW-Z (b, d) structure factor as a
function of $J/t$ and linear system size $L$ for $J_{z}/t=2$ and
$J_{z}/t=3$. There is no SDW-Z magnetic order when $J_{z}/t\le
2$ in the whole $J/t$ regime. Around the bulk quantum phase
transition critical point (QCP), SDW-XY structure factor shows a
power-law increasing tendency with system size $L$, however, the
power-law increasing exponent is less than 2 which means no SDW-XY
magnetic order will develop around the QCP in the thermodynamic
limit.}
  \label{fig:SzzStrFct}
\end{figure}

The Ising-like $J_{z}$ term in our Hamiltonian can be decompose
into the following three terms,
\begin{equation}
-\frac{J_{z}}{4}\sum_{i}\left[(\hat{n}_{1i\uparrow}-\hat{n}_{1i\downarrow})-(\hat{n}_{2i\uparrow}-\hat{n}_{2i\downarrow})\right]^{2}=-\frac{J_{z}}{4}\sum_{\xi,i,\sigma}\hat{n}_{\xi i\sigma}+\frac{J_{z}}{2}\sum_{\xi,i}\hat{n}_{\xi i\uparrow}\hat{n}_{\xi i\downarrow}+2J_{z}\sum_{i}\hat{S}_{1i}^{z}\hat{S}_{2i}^{z}
\end{equation}
The first term is the on-site potential term, the second term is
the on-site Coulomb repulsive interaction and the third term is
the Ising exchange interaction between two layer sites. When
$J_{z}\gg J$, $J_{z}$ will drive the system into a Ising
antiferromagnetic ordered (SDW-Z) state. We define the SDW-Z
antiferromagnetic magnetic order along $z$
direction as follows
\begin{equation}
\begin{split}
M_{AA}^{zz}(\mathbf{r}_{j}-\mathbf{r}_{i})=&S_{A_{1}A_{1}}^{zz}(\mathbf{r}_{j}-\mathbf{r}_{i})-S_{A_{1}A_{2}}^{zz}(\mathbf{r}_{j}-\mathbf{r}_{i})
-S_{A_{2}A_{1}}^{zz}(\mathbf{r}_{j}-\mathbf{r}_{i})+S_{A_{2}A_{2}}^{zz}(\mathbf{r}_{j}-\mathbf{r}_{i})\\
S_{mn}^{zz}(\mathbf{r}_{j}-\mathbf{r}_{i})=&\bra{\Psi}\hat{S}_{i}^{z}\hat{S}_{j}^{z}\ket{\Psi}/\braket{\Psi|\Psi},i\in m, j\in n
\end{split}
\end{equation}
From Fig.~\ref{fig:SzzStrFct}, there is no SDW-XY and SDW-Z
magnetic orders (and no time-reversal symmetry breaking) in the
whole $J/t>0$ parameter regime when $J_{z}/t\le 2.0$. However, when
$J_{z}/t=3.0$, SDW-Z order emerges in the middle of $J/t$ parameter
region.

\section{III. Energy curves}

\begin{figure}[htp!]
  \centering
  \includegraphics[width=0.7\columnwidth]{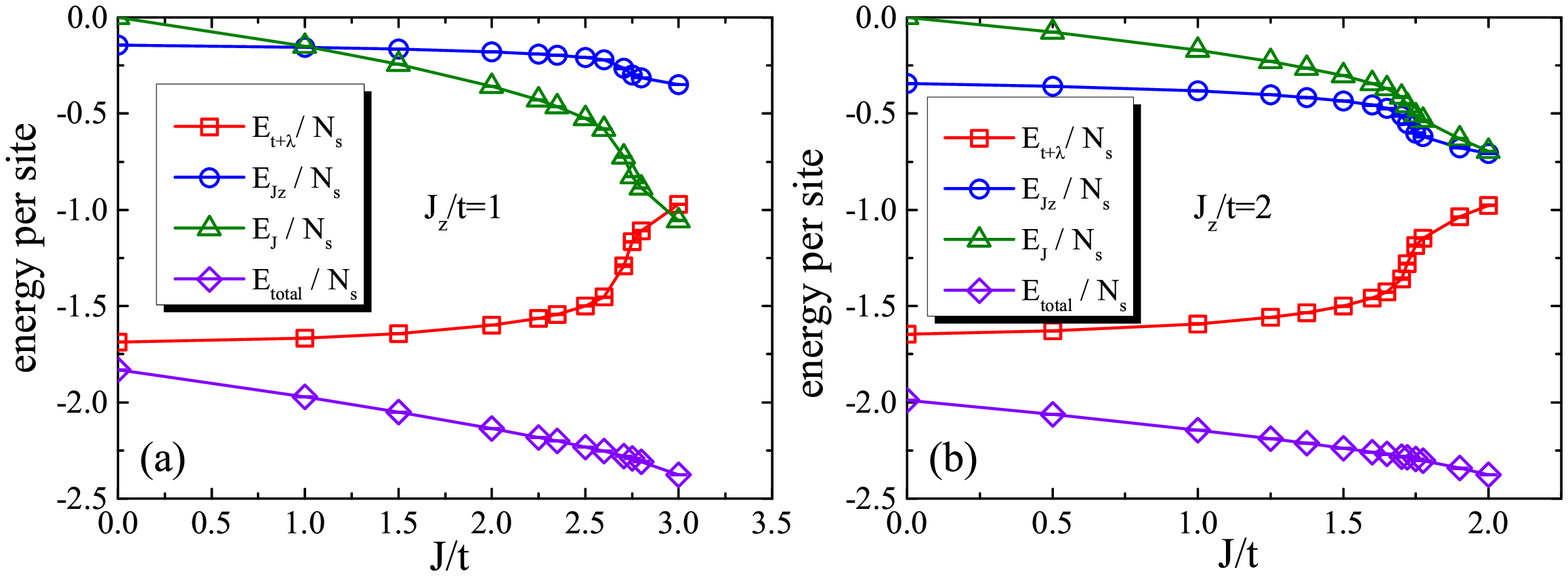}
\caption{Ground state energy per site as a function of $J/t$ when
$J_{z}/t=1$ and $2$. The linear system size used here is
$L=15$. Combined with Fig.~\ref{fig:SzzStrFct}, we can get the
phase diagram which is shown in Fig.1 (b) in the main text.}
  \label{fig:EngySite}
\end{figure}

We plot the expectation values of four parts of the Hamiltonian
in Fig.~\ref{fig:EngySite} as a function of $J/t$ for different
$J_{z}/t$ values. From the inflection point of the energy curves and
magnetic structure factor shown in Fig.~\ref{fig:SzzStrFct}, we
can obtain the approximate bulk quantum phase transition points
without calculating the energy gaps.

\section{IV. Other matrix elements of edge Green's function and O(4) correlation function}

In the main text, we only show the Green function between $A_{1}$
sublattice and $B_{1}$ sublattice in the same layer along the
ribbon edge, {\it{i.e.}}, an off-diagonal term of the edge Green's
function matrix. Here, we present that the diagonal parts of Green
function matrix also show similar behavior as the off-diagonal
part.

\begin{figure}[htp!]
  \centering
  \includegraphics[width=0.4\columnwidth]{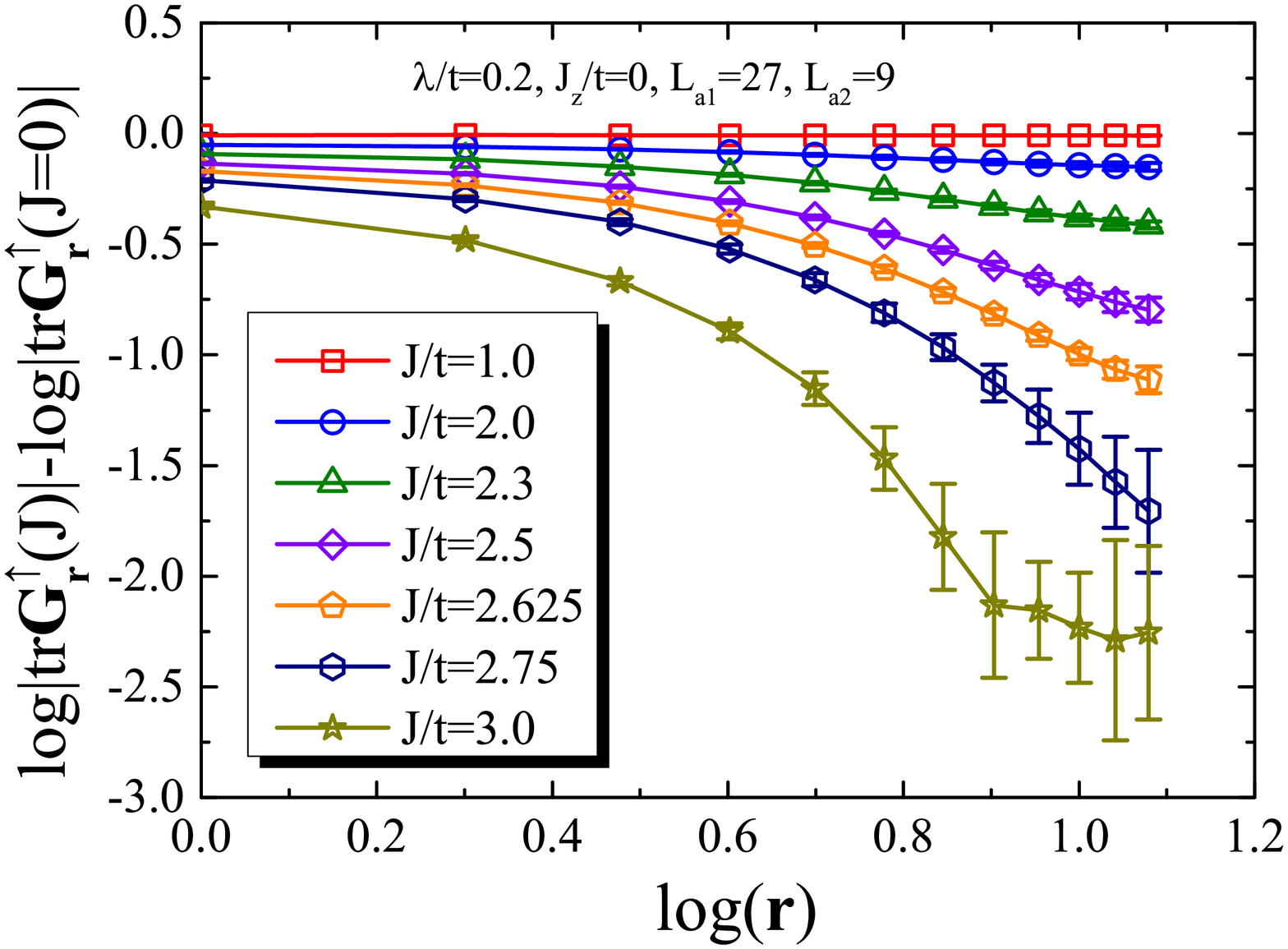}
\caption{The trace of single-particle Green's function matrix at
the ribbon edge as a function of $J/t$ when $J_{z}/t=0$.}
  \label{fig:TraceOccET}
\end{figure}

Fig.~\ref{fig:TraceOccET} shows the trace of single-particle
Green's function matrix
$\text{Tr}\mathbf{G}_{\mathbf{r}}^{\uparrow}=G_{A_{1}A_{1}}^{\uparrow}
+G_{A_{2}A_{2}}^{\uparrow}+G_{B_{1}B_{1}}^{\uparrow}+G_{B_{2}B_{2}}^{\uparrow}$
at the ribbon edge as a function of $J/t$ when $J_{z}/t=0$. The
diagonal part of single-particle Green's function at the edge also
shows the exponential decay before the bulk quantum phase
transition.

\begin{figure}[htp!]
  \centering
  \includegraphics[width=0.8\columnwidth]{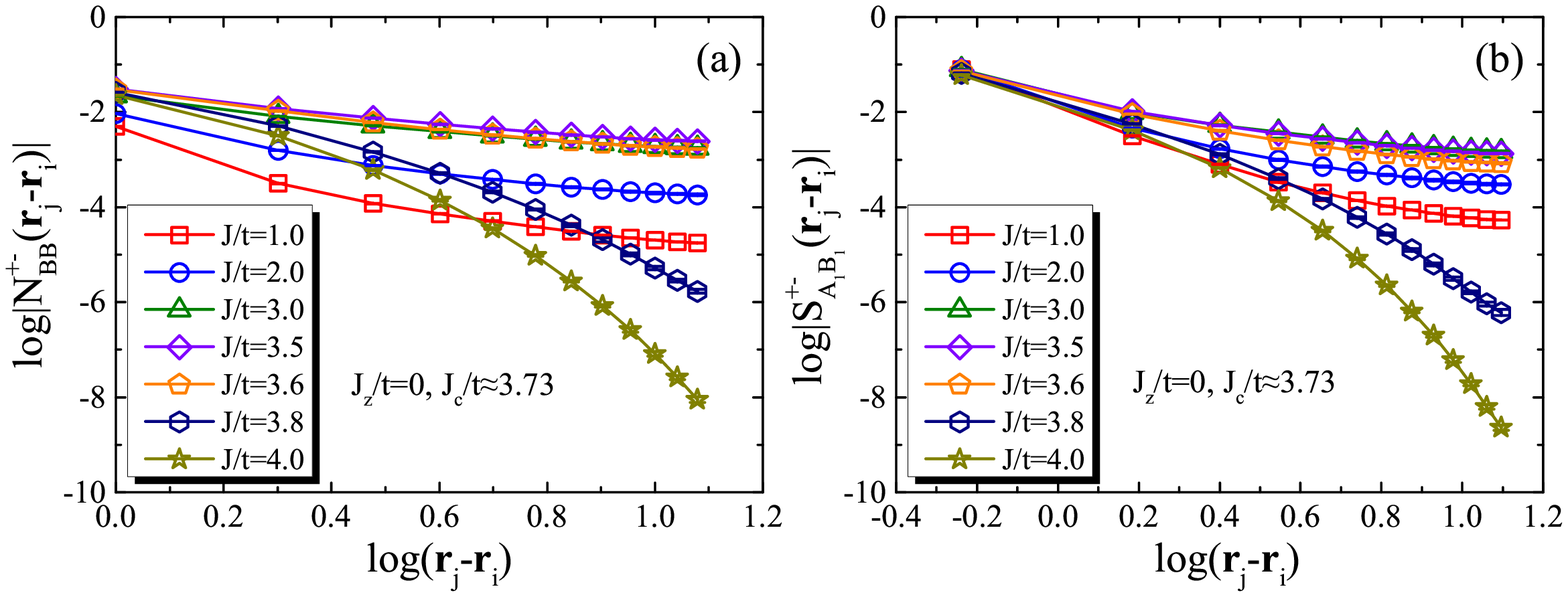}
\caption{The SDW-XY correlation functions $|N_{BB}^{+-}|$ and
$|S_{A_{1}B_{1}}^{+-}|$ at the ribbon edge as a function of $J/t$
when $J_{z}/t=0$.}
  \label{fig:OtherSDWXY}
\end{figure}

For the SDW-XY correlation matrix, we have show the
$|N_{AA}^{+-}|$ (with combined elements) in the main text. Here,
we also show you the power-law decay of $|N_{BB}^{+-}|$ and
$|S_{A_{1}B_{1}}^{+-}|$ before the bulk quantum phase transition
in Fig.~\ref{fig:OtherSDWXY}, where $N_{BB}^{+-}$ defines as
\begin{equation}
N_{BB}^{+-}(\mathbf{r}_{j}-\mathbf{r}_{i})=\frac{1}{2}[S_{B_{1}B_{1}}^{\pm}(\mathbf{r}_{j}-\mathbf{r}_{i})-S_{B_{1}B_{2}}^{\pm}(\mathbf{r}_{j}-\mathbf{r}_{i})
-S_{B_{2}B_{1}}^{\pm}(\mathbf{r}_{j}-\mathbf{r}_{i})+S_{B_{2}B_{2}}^{\pm}(\mathbf{r}_{j}-\mathbf{r}_{i})].
\end{equation}

\section{V. Finite-size effects}

\begin{figure}[htp!]
  \centering
  \includegraphics[width=0.8\columnwidth]{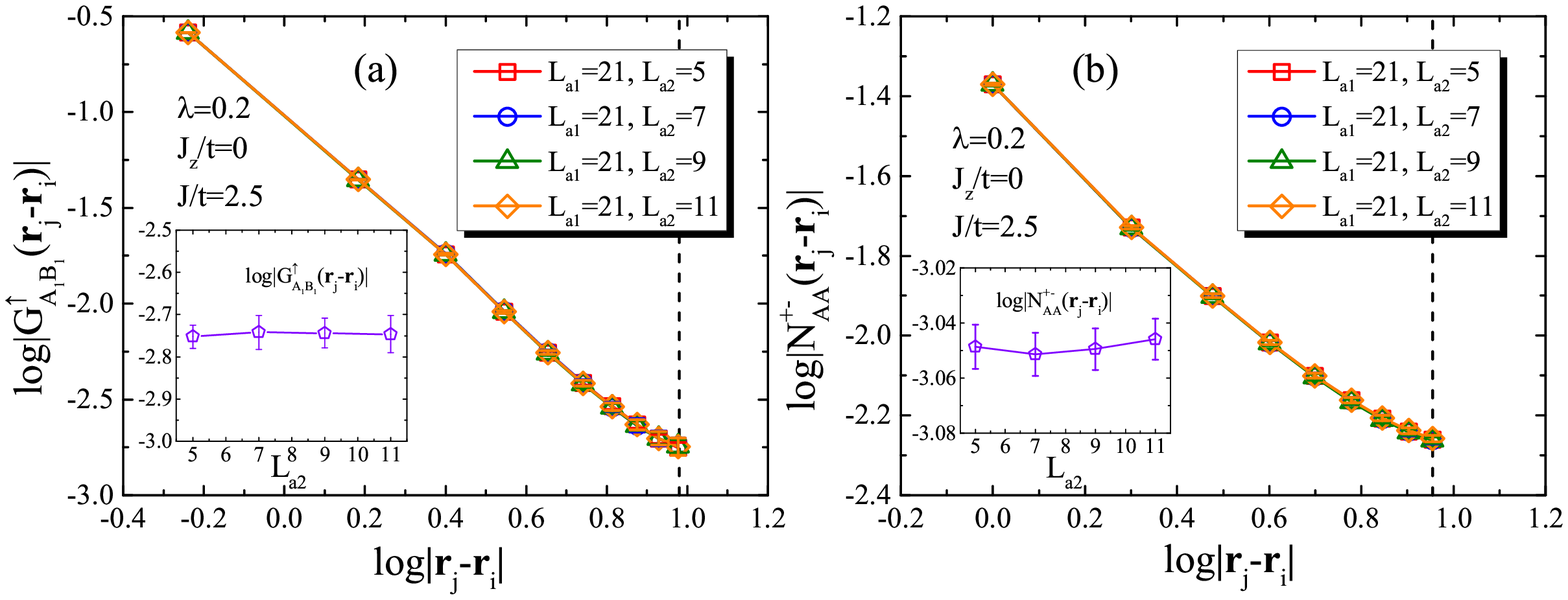}
\caption{The single-particle Green's function and SDW-XY
correlation function at the ribbon edge change little when we
increase $L_{a_{2}}$ from 5 to 11. The insets show the y-axis values of the right-most points as a function of ribbon width $L_{a_{2}}$.}
  \label{fig:La2FSE}
\end{figure}

In the main text, we mainly use the $L_{a_{1}}=27, L_{a_{2}}=9$
system size in the PQMC calculations. Here, we show that
$L_{a_{2}}=9$, which is the width of the ribbon, is large enough
to obtain thermodynamic limit behavior. As shown in
Fig.~\ref{fig:La2FSE}, when we increase the $L_{a_{2}}$ from 5 to
11, little change both in the single-particle Green's function as
well as two-particle bosonic correlation function, can be
observed.

\section{VI. Strange Correlator}

Apart from creating a physical spatial edge to study the edge
physics, we can also calculate the strange correlator to reflect
the physical edge between two topological distinct many-body
ground state wave functions~\cite{YZYou2014b,HQWu2015}.
\begin{equation}
C(r,r')=\frac{\bra{\Omega}\hat{\phi}(r)\hat{\phi}(r')\ket{\Psi}}{\braket{\Omega|\Psi}}
\end{equation}
we can define the single-particle strange correlator and spin
strange correlator by replacing the bra state with a topological
trivial state $\bra{\Omega}$ in Eq.~(4) in the main text. The
single-particle strange correlator also shows an exponential decay
before the bulk quantum phase transition while the spin strange
correlator remains power-law decay, indicating the interacting QSH
phase $\ket{\Psi}$ is topologically distinct from the trivial
phase $\bra{\Omega}$, and there exist gapless bosonic modes at the
spatial interface between two systems.

\begin{figure}[h]
  \centering
\includegraphics[width=0.8\columnwidth]{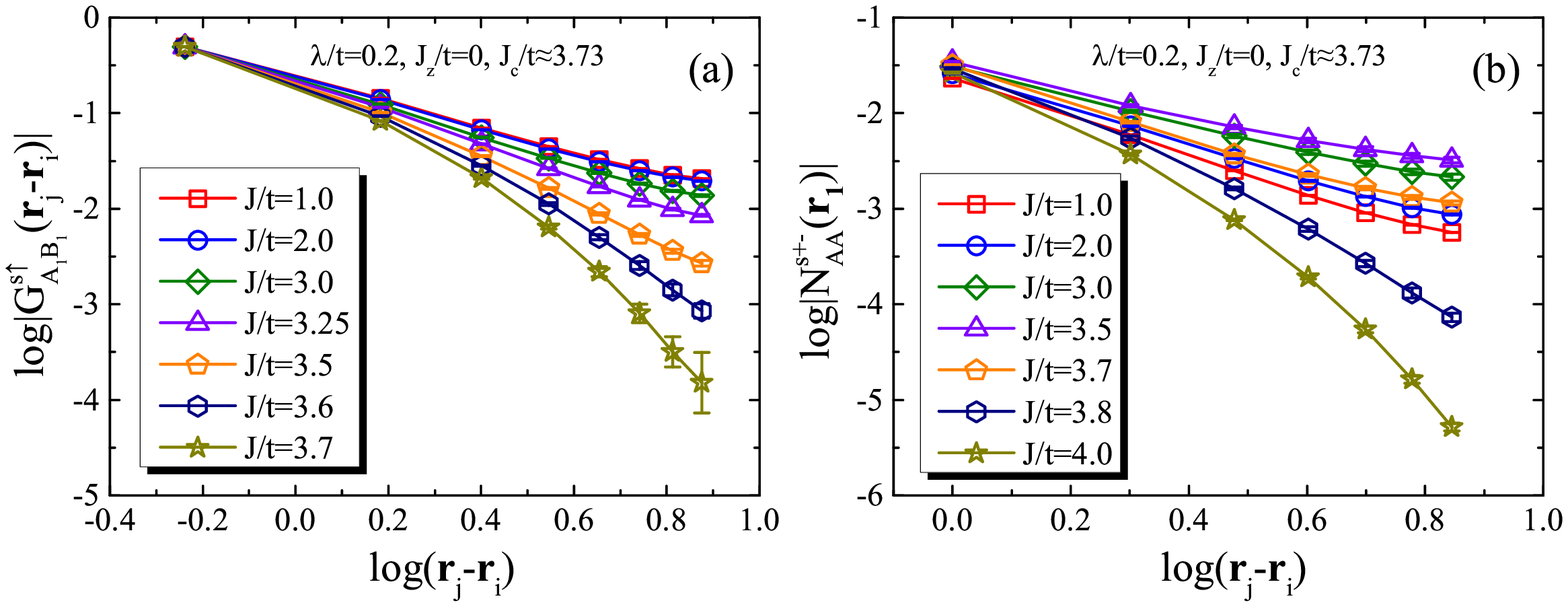}
\caption{single-particle strange correlator and SDW-XY strange
correlator as a function of $J/t$ when $J_{z}/t=0$.}
  \label{fig:StrCorr}
\end{figure}

\section{VII. on-site $U$ interaction}
\label{sec:UInt} 
\begin{figure}[h]
\centering
\includegraphics[width=0.8\columnwidth]{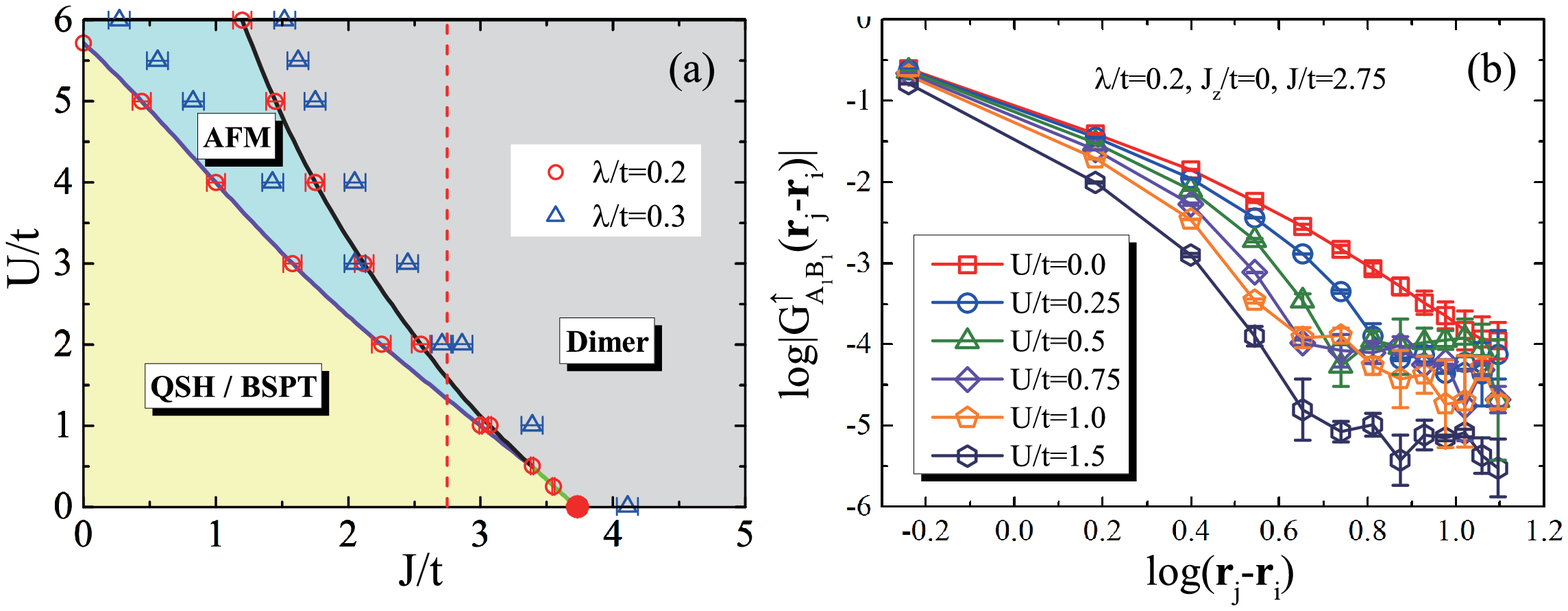}
\caption{(a) Phase Diagram of bilayer KMH model with on-site $U$
interaction and inter-layer $J$ interaction. The red line shows
the vertical phase path we used in Fig.~5 in
the main text. The exponential decay of single-particle Green's
function at the ribbon edge indicates that fermions are still
gapped when $U/t$ is increased at $J/t=2.75$.} \label{fig:UJPhase}
\end{figure}

The phase diagram of bilayer KMH model with
on-site
$U\sum_{i}(\hat{n}_{i\uparrow}+\hat{n}_{i\downarrow}-1)^{2}$
interaction and inter-layer $J$ interaction is shown in
Fig.~\ref{fig:UJPhase} (a). The phase boundaries are obtained from
the bosonic gap closing as well as the nonzero magnetic order
parameter in our previous paper Ref.~\cite{YYHe2016a}. Based on
the exponential decay of edge single-particle Green's function in
Fig.~\ref{fig:UJPhase} (b) and the power-law decay of edge SDW-XY
correlation function in Fig.~5 in the main text, we conclude that
the quantum spin Hall phase with \textit{finite} interaction $U$ and $J$
which is shown in Fig.~\ref{fig:UJPhase} (a) is also a bosonic SPT
phase.

\end{document}